**Title:** Annual net community production and carbon exports in the central Sargasso Sea from autonomous underwater glider observations

**Authors:** Ruth G. Curry[a,1], Michael W. Lomas[b], Megan R. Sullivan[a,2], Damian Grundle[c]

**Author Affiliations:**
[a] Bermuda Institute of Ocean Sciences / Arizona State University, St. Georges, Bermuda
[b] Bigelow Laboratory for Ocean Sciences, East Boothbay, ME, USA
[c] Arizona State University, Tempe, AZ USA

*Corresponding Author:* Ruth G. Curry
BIOS, 17 Biological Station, St. Georges, GE 01, Bermuda

**Email**: ruth.curry@bios.asu.edu
mlomas@bigelow.org
megan.sullivan1@uri.edu
damian.grundle@bios.asu.edu

[1] 3 Wright Street, Falmouth, MA 02540 USA
[2] Graduate School of Oceanography, University of Rhode Island, 215 S. Ferry Rd., Narragansett, RI USA

# Annual net community production and carbon exports in the central Sargasso Sea from autonomous underwater glider observations

**Abstract**
Despite decades of ship-based observations at the Bermuda Atlantic Timeseries Study (BATS) site, ambiguities linger in our understanding of the region's annual carbon cycle. Difficulties reconciling geochemical estimates of annual net community production (ANCP) with direct measurements of nutrient delivery and carbon exports (EP) have implied either an insufficient understanding of these processes, and/or that they are playing out on shorter time and spatial scales than resolved by monthly sampling. We address the latter concern using autonomous underwater gliders equipped with biogeochemical sensors to quantify ANCP from mass balances of oxygen ($O_2$) and nitrate ($NO_3^-$) over a full annual cycle. The timing, amplitude and distributions of $O_2$ production, consumption, and $NO_3^-$ fluxes reaffirm ideas about strong seasonality in physical forcing and trophic structure creating a dual system: i.e. production fueled by $NO_3^-$ supplied to the photic zone from deeper layers in the first half of the year, versus being recycled within the upper ocean during the second half. The evidence also supports recently proposed hypotheses regarding the production and recycling of carbon with non-Redfield characteristics, depleted in nitrogen and phosphorus, to explain observed patterns of high NCP in the absence of significant $NO_3^-$ supply. It further identifies significant contributions to ANCP and EP potentially linked to vertically migrating communities of salps in spring after all convective activity has ceased. The improved resolution of the datasets, combined with more precise definitions of photic and subphotic integration depths, brings the estimates of ANCP and EP into better alignment with each other.

## 1. Introduction

When integrated over an annual cycle, the net biological production of organic matter in the upper ocean is equal to the annual net community production (ANCP) defined as the difference between the rate of net primary production (NPP) and respiration by heterotrophs. ANCP is often estimated using the mass balance of metabolites, e.g., dissolved oxygen ($O_2$) or dissolved inorganic carbon (DIC) inventories, and restricted to periods when active vertical mixing has stopped (Emerson, 2014 and references therein). ANCP is conceptually equivalent to the export production (EP) of organic carbon (dissolved and particulate)—the carbon flux that is transported from the surface ocean to deeper waters and sediments via the "biological pump". To this end, it is quantitively equivalent to new production (NP) that is supported by externally derived nitrogen sources, as opposed to the nitrogen that is recycled within the euphotic zone (Dugdale & Goering, 1967).

Understanding both the geographic distribution and biological mechanisms governing ANCP and EP is important for several reasons. These include expectations that: 1) regional changes in ANCP and EP may alter the efficiency of carbon export to depth and the geographic storage of those exports (Kwon et al., 2009); 2) changes in the depth of organic matter remineralization can also directly affect levels of atmospheric $CO_2$ and thermocline $O_2$ distributions (Kwon et al., 2009; Tanioka et al., 2021); and 3) non-Redfield behavior in phytoplankton may counter EP changes projected to result from future warming and stratification (Lomas et al. 2022; Sullivan et al. 2024). In short, this is a complex system with significant implications for marine biogeochemical budgets, yet only a few sampling programs in the ocean provide the resolution and longevity to estimate and constrain ANCP.

Comparing two long-term oceanographic time-series stations – the Hawaii Ocean Times-series (HOT) in the North Pacific and the Bermuda Atlantic Time-series Study (BATS) in the central Sargasso Sea – Brix et al. (2006) suggested that differences in their NCP, EP and NP characteristics reflected local environmental conditions, particularly nutrient input, and its impact on ecosystem structure. Whereas NCP at HOT was fueled by a microbially-dominated "regeneration loop" characterized by a low EP:NPP ratio, they postulated that at BATS, strong physical forcing and elevated nutrient delivery supported a seasonally active "export pathway" (high EP: NPP ratio) from February to May, which reverted to a regeneration loop in the summer and fall months. This seasonality was expressed as hysteresis in the relationship between NPP and NCP over the annual cycle that was not present at HOT. They were, however, unable to establish a predictive model based on physical forcing to support their hypothesis.

At BATS, geochemical approaches using annual mass balances of $O_2$ and DIC yield comparable estimates of ANCP —i.e., 45.6 ± 14.4 g C m$^{-2}$ y$^{-1}$ (Jenkins and Goldman, 1985; Spitzer and Jenkins, 1989; Jenkins and Doney, 2003, Gruber et al., 1998; Stanley et al., 2011; Emerson, 2014). The magnitude of the error bars reflects uncertainties related to both temporal (monthly) sampling frequencies and spatial heterogeneity associated with the active regional eddy field. Total annual primary production at BATS has been estimated to be in the range of 114 to 215 g C m$^{-2}$ (Steinberg et al., 2001; Lomas et al., 2013). Based on the traditional NP paradigm (i.e. no euphotic zone nitrification) and using the canonical view of how the fraction of total primary production fueled by $NO_3^-$ varies with changes in total primary production (i.e. *f*-ratio, Eppley and Peterson, 1979), Lipschultz et al. (2002) equated this to an *f*-ratio of ~0.3. Applying this value to the total annual primary production range results in an annual NP estimate of 34-65 g C m$^{-2}$ y$^{-1}$, which aligns with the above-cited geochemical ANCP estimates. Assuming Redfield rationale, these carbon-based ANCP estimates are equivalent to 0.52±0.2 mol N m$^{-2}$ y$^{-1}$, which compares favorably with N-based ANCP estimated from $O_2$ mass balance and independent tracers ($^3$He), i.e., 0.6 ±0.2 mol N m$^{-2}$ y$^{-1}$ (Jenkins, 1988). The sum of these studies thus implies a convergence in ANCP derived from geochemical methods and the amount of $NO_3^-$ required to support it. This convergence is further supported by natural abundance isotopic analyses of $NO_3^-$ in the Sargasso Sea that suggest all $NO_3^-$ in the euphotic zone is transported upwards from deeper waters (Fawcett et al., 2011) as opposed to arising from *in situ* nitrification (i.e. regenerated nitrate within the euphotic zone, e.g., Ward 2005; Yool et al. 2007; Grundle and Juniper 2011; Grundle et al. 2013). However, these geochemical evaluations of ANCP, EP and NP have been persistently at odds with direct measurements of the same (e.g., carbon exports and nutrient supply) resulting in budget imbalances that have defied reconciliation for several decades. The essential issues underlying these imbalances have been succinctly described by Lomas et al (2013),

Emerson (2014), and recently revisited by Fawcett et al. (2018), thus we only briefly summarize them here.

First, directly measured carbon exports, i.e., via sinking particles (POC), zooplankton migration ($F_Z$), and dissolved organic carbon (DOC), account for only a fraction of NCP estimated from $O_2$ production and DIC drawdown. POC export has been estimated as $10.5\pm1.7$ g C m$^{-2}$ y$^{-1}$ (Lomas et al., 2013), $F_Z$ is equivalent to $1.4\pm0.5$ g C m$^{-2}$ y$^{-1}$ (Steinberg et al., 2012), while the DOC portion contributes $10.2\pm6.6$ g C m$^{-2}$ y$^{-1}$ (Carlson et al., 1994; Hansell and Carlson, 2001). These sum to $22.1\pm8.8$ g C m$^{-2}$ y$^{-1}$ which is approximately half of geochemical ANCP estimates. The methods underpinning these measurements all have caveats (Fawcett et al., 2018 and references therein). For example, several issues thought to contribute to under sampling by surface-tethered sediment traps have been identified, yet, neutrally buoyant sediment traps, free from most hydrodynamic biases, yield similarly low EP values (Stanley et al. 2004, Owens et al., 2013).

Secondly, the drawdown of DIC and high NCP measured in the summertime near-surface layers (Gruber et al., 1998) appears to occur in the absence of a circulation-based mechanism of nutrient delivery. Fawcett et al. (2018) provided an extensive description of possible mechanisms supplying the missing nutrient transport but argued effectively in favor of a non-Redfield explanation: i.e., production and export of nutrient-poor organic matter by phytoplankton adapted to low-nutrient conditions. They hypothesized that this export of low nutrient DOM occurs in the form of gel-like organic matter (GLOM), rich in carbon but poor in nitrogen and phosphorus, which sinks into the shallow subsurface layers where it is respired by heterotrophic bacteria. They noted that this would explain both the vertical separation between $O_2$ production in the near-surface waters, in the absence of a mechanical nutrient supply, and an observed decrease in $O_2$ concentration without an increase in nitrogen in the deeper subsurface layers. They further suggested that the physical properties of GLOM could preclude its capture by sediment traps thus explaining the lack of prior detection.

In light of the challenges to closing biogeochemical carbon budgets at BATS, despite nearly four decades of monthly observations, it has been suggested that either we do not understand the nutrient delivery processes as well as we thought, that there are processes we are missing, and/or that they are playing out on shorter time and spatial scales than can be resolved by monthly ship-based sampling. Possible explanations include non-measured ecosystem mechanisms (e.g., diel vertical migrators such as salps producing large fecal pellets that are not captured efficiently by sediment traps), changes in trophic structure that can lead to more efficient nutrient recycling (Tanioka et al. 2021; Russo et al. 2024), and/or changes in the stoichiometric composition of the source material (i.e., phytoplankton) that contributes to enhanced nutrient recycling (Lomas et al. 2021, 2022). A corollary to these is that the processes of production and respiration may be occurring on different timescales (e.g., Emerson et al. 2002, McAndrew et al. 2007). The latter can now be addressed with autonomous underwater vehicles and new biogeochemical sensors which provide the endurance, temporal resolution and accuracy required to resolve processes on shorter timescales (hours to days) associated with storms, ocean fronts and phytoplankton growth (Rudnick, 2016). As evidence of this, glider- and profiling float-based budgets of $O_2$, $NO_3^-$, chlorophyll fluorescence, and POC have been applied to estimate NCP and EP at HOT (Nicholson et al. 2008; Nicholson et al. 2015) and in the subpolar western North Atlantic (Alkire et al. 2012; Alkire et al. 2014). Those studies quantified biogeochemical processes over periods of months and established the applicability of these technologies to acquiring datasets in remote regions, over extended stretches of time, and in economically feasible ways.

Here we combine upper ocean properties measured by underwater autonomous gliders near the BATS site with inventories of phytoplankton biomass derived from flow cytometry measurements (FCM), to explore relationships among NCP, NP, and EP over a full annual cycle. The enhanced resolution afforded by these glider observations provides insights that are both novel and complementary to previous efforts to quantify carbon production and export. Acquired in 2017 and 2018, we have cast these observations into a framework defined by seasons and vertical layers chosen specifically to reflect dynamical, physical and biogeochemical boundaries in the Sargasso Sea, and we use it to quantify the timing, amplitude and vertical distributions of NCP, nutrient fluxes, and the phytoplankton community structure underpinning them. The gliders' high temporal resolution (8-16 profiles per day combined into daily averages) enabled a detailed accounting of these for both the photic and subphotic layers, while our layer definitions more closely tracked day-to-day variability in their vertical structure compared to integrating over fixed depth intervals. The glider-based estimates confirmed strong seasonality in NCP and $NO_3^-$ fluxes and identified periods of $O_2$ production and consumption associated with distinct driving mechanisms over the course of a year. The evidence reinforces and supplements the findings of Fawcett et al. (2018) regarding the production and recycling of carbon as low-nutrient organic matter with non-Redfield characteristics. It also affirms ideas about seasonality in the export pathway vs. regeneration loop proposed by Brix et al. (2006) and offers new insight as to why physical forcing alone did not lead to predictability in their study. Finally, the improved resolution of the data combined with more precise definitions of photic and subphotic integration depths brings the estimates of ANCP and EP into better alignment.

We have organized and presented these results as follows. In section 2, the physical framework is defined, while the glider and phytoplankton observational datasets are described in Section 3. In section 4, biogeochemical fluxes of $O_2$ and $NO_3^-$ are derived and applied to estimate cumulative NCP in layers over the seasonal cycle. The gliders' optical chlorophyll and backscatter profiles are then scaled to phytoplankton biomass and community structure determined from FCM providing perspectives on the vertical distribution and timing of biological productivity, and their relationships to vertical nutrient delivery within the photic zone and remineralization below it. These are discussed in the context of the present understanding of carbon cycle processes derived from the BATS program. Conclusions, unresolved questions, and future ways to address them are put forward in section 5.

 2. The physical framework

The physical framework we have developed partitions the ocean at BATS into ten vertical zones ($VZ_{0-9}$) and four seasons (*Mixed*, *Spring*, *Stratified* and *Fall*), defined in Tables 1 and 2. The seasons are referred to by capitalized and italicized labels. Vertical bounds are determined from depth, density and chlorophyll criteria (commonly measured, for example, by a CTD and fluorometer), while seasonal boundaries are set by the first emergence and disappearance of the chlorophyll maximum (CM) layer into the surface mixed layer (ML) over each annual cycle. The specific start and end dates for these boundaries vary from year to year, but these season definitions more closely reflect the interplay of locally varying physical and biogeochemical processes compared to conventional winter/spring/summer/fall designations in ways that facilitate interannual comparisons. As visual evidence of this, time series of daily-averaged properties from 1.5 years of glider observations depict the boundaries between vertical zones, the timespan of seasons, and their relationships to biogeochemical properties: stratification, oxygen anomaly ($O_2 - O_{sat}$), optical backscatter measured at a wavelength of 700 nm ($BB_{700}$), chlorophyll fluorescence (ChlF), and $NO_3^-$ concentration (Fig. 1).

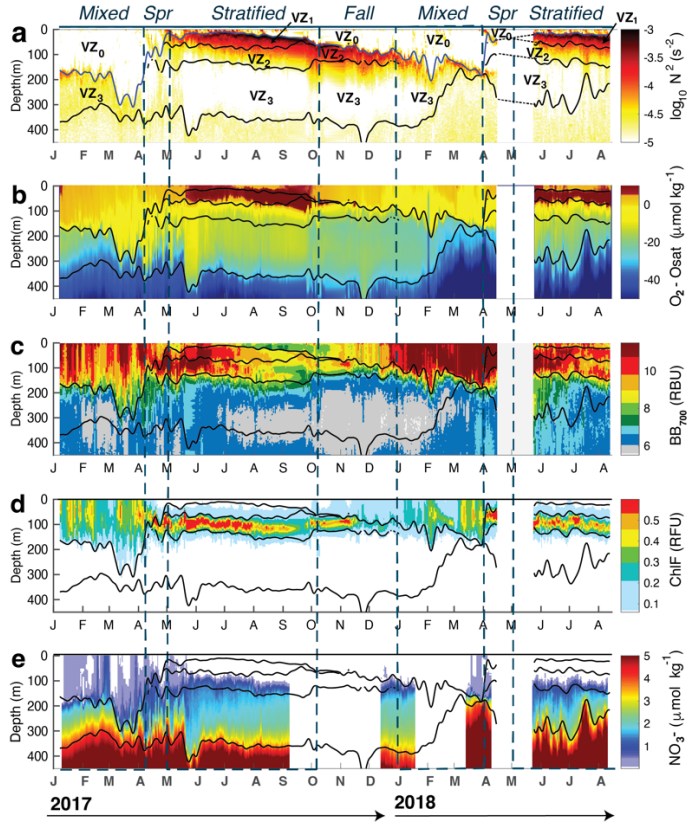

*Figure 1* **Time series of upper ocean properties (0-450m) measured by gliders from Jan 2017 to Aug 2018 with the physical framework season and layer definitions**. (**a**) buoyancy frequency ($\log_{10} N^2$, an indicator of vertical stratification), (**b**) $O_2$ anomaly, defined as dissolved oxygen minus the $O_2$ saturation concentration, (**c**) optical backscatter, (**d**) chlorophyll fluorescence, and (**e**) $NO_3$ (sampled on a limited number of missions). Months are labeled along the x-axes. Heavy contours depict the boundaries of vertical zones 0 – 3, labeled in **a**. Note that $VZ_1$ and $VZ_2$ appear seasonally – they are subsumed into $VZ_0$ during the *Fall* and *Mixed* seasons respectively. Vertical dashed lines depict season boundaries, defined in Table 2 and labeled above the top panel. A gap in the time series from April – May 2018 reflects equipment failure.

**Table 1.** Vertical Zone Definitions

| # | Name | Definition |
|---|---|---|
| $VZ_0$ | Surface Mixed Layer (ML) | 0 m to mixed layer depth (MLD, defined as the depth where density = sea surface $\sigma_\theta$ + 0.125 kg m$^{-3}$ |
| $VZ_1$ | Upper Stratified Euphotic Zone (USEZ) | MLD to top of CM layer |
| $VZ_2$ | Chlorophyll Maximum (CM) Layer | Portion of chlorF profile >= 0.35 * maximum chlorF |
| $VZ_3$ | Winter Mode Water (WMW) Layer | Base of CM layer to $\sigma_\theta$ =26.32 |
| $VZ_4$ | Ventilated Thermocline Layer (VTL) #4 | $\sigma_\theta$ = 26.32 to 26.50 |
| $VZ_5$ | VTL #5 | $\sigma_\theta$ = 26.50 to 26.70 |
| $VZ_6$ | VTL #6 | $\sigma_\theta$ = 26.70 to 26.90 |
| $VZ_7$ | VTL #7 | $\sigma_\theta$ = 26.90 to 27.10 |
| $VZ_8$ | VTL #8    Deep O2 minimum | $\sigma_\theta$ = 27.10 to 27.30 |
| $VZ_9$ | VTL #9 | $\sigma_\theta$ = 27.30 to 27.50 |

**Table 2.** Season Definitions

| Season | Begins |
|---|---|
| *Mixed* | MLD is deeper than DCM |
| *Spring* | MLD abruptly shoals and CM layer emerges below it |
| *Stratified* | MLD is consistently shallower than top of CM layer |
| *Fall* | First entrainment of top of CM layer into ML; DCM remains deeper than MLD |

2.1. Upper Ocean

The upper four layers, which respond strongly to local atmospheric forcing, begin with the surface ML, $VZ_0$, and extend down to the base of $VZ_3$, the local winter mode water (WMW) layer. The latter is defined by a density ($\sigma_0$ = 26.32 kg m$^{-3}$) reflecting the densest winter ML formed in the local BATS climatological record. Two intervening layers, $VZ_1$, the upper stratified euphotic zone (USEZ), and $VZ_2$, the CM layer, appear seasonally from *Spring* to *Fall*, but are entirely absent throughout the *Mixed* season. The depth and thickness of these layers can vary greatly over the annual cycle as a consequence of physical processes (e.g. convective mixing, solar heating, light penetration, and the presence of mesoscale eddies) which in turn can elicit strong biogeochemical responses that may include changes in concentrations of plankton, dissolved organic carbon, oxygen, nutrients, and chlorophyll (e.g. McGillicuddy, 2016).

Our framework makes use of three ML depth (MLD) variables which reflect the impact of slightly differing mixing dynamics on the density profile. 1) $ML_{dens125}$ is defined by a density threshold criterion as the depth where $\sigma_0$ exceeds the surface density by 0.125 kg m$^{-3}$. It generally reflects the deepest reach of seasonal convective mixing, exhibits the lowest frequency variability over the annual cycle, and is the deepest of the three MLD variables. 2) $ML_{densT2}$ is defined after Sprintall and Tomczak (1992) as the depth where $\sigma_0$ exceeds the surface density + 0.2 $\alpha$, where $\alpha$ is the thermal expansion coefficient. It often marks intermediate episodes of convective mixing which do not reach to $ML_{dens125}$. 3) $ML_{bvfrq}$ is defined (Diaz et al., 2021) as the depth where buoyancy frequency ($N^2$) first exceeds the standard deviation of $N^2$ in the water column. It responds to diurnal scales of active mixing, exhibits the highest frequency variability and shallowest depths of the three MLD variables. To illustrate their variability and inter-relationships, all three MLDs have been compared both as daily average values over the annual cycle and on shorter timescales (a few hours) for a subset of the year (Fig. S2). All MLs are bounded at their base by enhanced density gradients of some degree, but the strongest barrier layer occurs at the base of $ML_{dens125}$ in the *Stratified* season (Fig.1a). The three MLDs may be coincident or divergent, depending on mixing intensity and the relaxation of horizontal buoyancy gradients on relatively short time scales (hours to days). They provide a basis for diagnosing stratification dynamics and relating them to other phenomena observed in the water column.

In this framework, the base of $VZ_0$ is defined by $ML_{dens125}$. The layer is weakly stratified, but its vertical extent varies greatly over the annual cycle (Fig. 1a). $VZ_1$ and $VZ_2$ emerge seasonally in *Spring*, become increasingly stratified from the surface downward during the summer months, and are progressively eroded away by convective mixing in *Fall*. The upper and lower bounds of $VZ_2$, are defined as the depths where ChlF is equal to 0.35 of the maximum value in the profile – i.e. above and below the depth of the chlorophyll maximum (DCM), (Figs. **1** and **S1**). Here we differentiate between the terms "DCM" and "CM layer" with the former being a point, and the latter occupying a depth range above and below it. $VZ_1$ occupies the water column between the MLD and top of the CM layer. The $VZ_2$ / $VZ_3$ boundary corresponds to the base of the euphotic zone, and the upper limits of measurable $NO_3^-$ concentrations. Light levels at the base of $VZ_2$ measured by a PAR sensor deployed on multiple glider missions correspond to a mean value equal to 0.20% of the surface radiance with a standard deviation of 0.15%, compared to 1.30% ± 0.81% at the DCM itself. $VZ_1$ which is largely devoid of $NO_3^-$, is characterized by low ChlF but high $BB_{700}$, and exhibits super-saturated $O_2$ concentrations during the *Stratified* season. $VZ_2$ exhibits measurable (but low) concentrations of $NO_3^-$, high ChlF and high $BB_{700}$ concentrations, and a vertically decreasing gradient of $O_2$.

The WMW layer ($VZ_3$) is very weakly stratified, largely devoid of ChlF, and relatively low in $NO_3^-$ compared to deeper layers. Its characteristics overlap with the Subtropical Mode Water (STMW) which is seasonally formed by convective mixing some distance to the northeast in the subtropical gyre (Joyce 2012), but we use WMW here to distinguish its slight differences from the definition of STMW. The top of $VZ_3$ is characterized by a weak gradient in $NO_3^-$, while its base marks the top of a sharp nitracline, below which $NO_3^-$ concentrations exceed 4 µmoles $kg^{-1}$.

2.2. Deeper Layers

Below $VZ_3$, the water column is arbitrarily partitioned by potential density layers (in increments of 0.2 kg $m^{-3}$) reflecting the dominance of flow along isopycnals beneath the local wintertime MLD (Table 1, Fig. 2). These layers are designated ventilated thermocline layers (VTL) #4-9. The mean gyre circulation slowly moves water in a southwestward direction past Bermuda at a rate of ~4 km $day^{-1}$ from locations generally to the east where they are exposed to atmospheric conditions (i.e. ventilated) during the winter months, then subducted beneath the surface and advected within the anticyclonic gyre (Luyten et al., 1982). Certain physical characteristics (e.g. stratification, temperature, salinity, and $O_2$ concentration) are set at the time and location of this air-sea interaction, although some of these undergo modification along their subsequent flow pathways. The "age" of waters increases with depth and density in the main thermocline, reflecting the longer time and distance traveled since they were ventilated. This is reflected in generally decreasing $O_2$ concentration– or increasing apparent oxygen utilization (AOU) – with depth as a consequence of biological respiration. The waters in VTL#8 are associated with the deep $O_2$ minimum of the subtropical main thermocline and are thus the "oldest" having been ventilated on the eastern side of the gyre, while waters with density greater than $\sigma_\theta$ = 27.20 are inferred by dynamical constraints to have originated in a shadow zone of stagnant flows on the gyre's eastern boundary (Luyten et al., 1982).

Variability in vertical structure of these layers is strongly influenced by the gyre's mesoscale eddy field, which manifests as a rise and fall of layer depths on timescales of order 1-2 months and spatial scales of order 100-200 km (Fig. 2). Daily average geopotential anomaly (GPA) computed from glider density profiles and compared to a monthly reference GPA curve computed from the BATS time series reveals two types of eddies transiting the region between Oct 2017 and April 2018, and their contrasting impacts on the depths of vertical zones from $VZ_3$ downward. From October to January, a series of Mode Water eddies (MWE) characterized by a thickening of $VZ_3$ and deepening of layers beneath it, as well as by higher-than-normal GPA, occupied the sampling region. From February to April 2018, a cyclonic eddy (CE) characterized by low GPA resulted in uplifted layers beneath $VZ_0$. Eddies not only affect lateral transports of physical properties and biological communities through the Sargasso Sea, but also modulate the proximity of nutrient-rich waters to the euphotic zone which drives spatial and temporal variability in biological carbon pump processes, including NCP and EP (McGillicuddy et al., 2007, Benitez-Nelson and McGillicuddy, 2008; McGillicuddy, 2016).

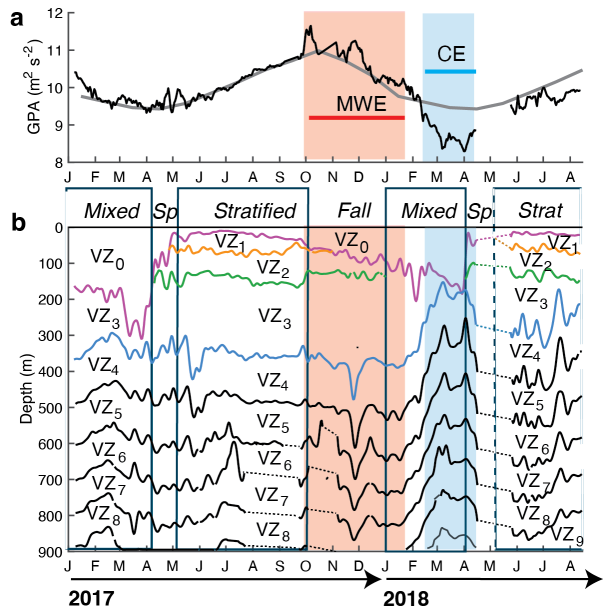

*Figure 2.* **Hydrographic structure in 2017-2018**. (**a**) Geopotential anomaly (GPA) at the sea surface relative to 500 db (daily averages, black line) derived from glider density profiles. The gray curve depicts the seasonal cycle of monthly averaged GPA derived from the BATS observations. Significant departures from the monthly average are the signature of eddies: e.g. a series of Mode Water Eddies (MWE, red bar) and a Cyclonic Eddy (CE, blue bar). (**b**) Depth of vertical zone boundaries ($VZ_0$ through $VZ_9$) over the 18-month glider time series. The bases of the top 4 layers are differentiated by line color for clarity. Season boundaries are depicted by vertical lines and labeled above the top axis. Colored boxes depict the presence of mesoscale eddies at the sampling site to highlight their effect on vertical water column structure.

2.3. Seasons

Because conventional season designations (winter, spring, summer, fall) are neither geographically uniform nor correspond strictly to calendar months, our framework divides the annual cycle into *Mixed* and *Stratified* seasons with intervening *Spring* and *Fall* transitions, whose temporal boundaries are determined by the relative depths of $ML_{dens125}$ and the CM layer (Table 2). The *Mixed* season begins when $VZ_0$ first punches through the base of $VZ_2$ (i.e. when the DCM is shallower than the MLD) and ends when the base of $VZ_0$ rapidly shoals, and the CM layer first emerges, usually sometime in late March or early April. The details and timing vary interannually, but this boundary begins the start of the *Spring* season, which persists until episodic mixing no longer penetrates below the top of $VZ_2$. This transition is frequently short (a few weeks), leading to the *Stratified* season during which the MLD remains consistently shallower than the top of $VZ_2$, and $VZ_1$ is present, typically from June to October. The *Fall* transition begins when $VZ_1$ is mixed away but $VZ_0$ is shallower than the DCM and it ends when convective mixing causes $VZ_2$ to disappear for the remainder of the year.

While high frequency sampling (e.g. producing daily average density and chlorophyll profiles) enables the precise identification of the start and end of seasons, season boundaries can readily be discerned from more granular datasets, e.g. the BATS record of monthly sampling, once the depth boundaries of $VZ_0$ and $VZ_2$ have been determined. By definition, the progression of seasons is monotonic – e.g. *Spring* does not revert to *Mixed* even if the CM layer is incorporated into $VZ_0$ by an episode of storminess, and *Fall* does not lapse back into the *Stratified* season after the MLD first begins to erode the top of the CM layer, even if $VZ_1$ reappears.

### 3. Data and Methods
3.1. Glider observations.

Between February 2017 and September 2018, time series of physical and biogeochemical properties were acquired near the BATS site (31° 40'N, 64° 10'W) using three separate Slocum G2 gliders deployed in 10 consecutive missions. Each glider carried a science payload that included a pumped Seabird CTD, a WetLabs ECOpuck model FLBBSLC measuring ChlF and $BB_{700}$, and an Aanderaa optode model 4831 measuring dissolved $O_2$. The vehicle was piloted to spiral around a 0.5 km box (essentially holding station) and profile between 0 and ~900 meters depth sampling in both up/down directions and completing 7-8 dives in a 24-hour period. For five missions a glider was additionally equipped with a Submersible Underwater Nitrate Analyzer (SUNA) that profiled on the climb cycle of every dive.

Post-mission corrections and quality assessments were undertaken for each individual science sensor following best practice procedures described by Woo and Gourcuff (2021). Monthly ship-based CTD and water sample profiles were acquired on BATS cruises within 10 km of the glider and used to calibrate the salinity and $O_2$ profiles to correct them for offsets in each mission. Prior to each deployment, a 2-point lab calibration (100% and 0% $O_2$ concentrations) was applied to the optode following manufacturer guidelines (AADI, 2014). In addition to lab and *in situ* calibrations, the optode data were corrected for the following: 1) physical distance between the glider CTD and optode mount; 2) sensor response time (following Bittig and Kortzinger, 2018); and 3) temperature, salinity and pressure effects using Stern, Volmer and Uchida formulas (e.g., Uchida et al., 2008). For the SUNA sensor, a pre-mission calibration was conducted in the lab using deionized water and manufacturer software (Seabird Scientific, 2017). Post-mission, another lab calibration was performed, and the SUNA data were corrected for pressure, temperature and salinity following the algorithm of Sakamoto et al. (2009), additionally accounting for the distance from the instrument's optical window to the glider CTD. The sensor exhibited a drift over the course of each mission, the rate of which was estimated by linear interpolation to offsets from monthly BATS profiles at an isopycnal level deeper than 400 m. ChlF profiles were corrected for quenching by determining the depth at which consecutive day/night profiles diverged, then replacing the daytime values with nighttime values above that depth (Thomalla et al., 2018). Drift in the sensor response over the course of each mission was estimated for the fluorometer from a linear fit to the difference from zero on an isopycnal level near 350 m depth. The missions were inter-calibrated by comparing profiles acquired near each other in space and time, as a function of density.

Daily average profiles were computed for all measured and derived properties from profiles spanning the time between local sunrise on consecutive days. GPA at the sea surface relative to 500 db was evaluated from daily-averaged density measured by the gliders and compared to a monthly average GPA curve derived from the BATS dataset as a diagnostic of mesoscale eddies. The years 2010 to 2020 were used for constructing the average to reflect recent warming of the upper ocean (Bates and Johnson, 2020) which significantly affects GPA. Daily inventories of $O_2$ (mol $O_2$ m$^{-2}$), $NO_3^-$ (mol $NO_3^-$ m$^{-2}$), ChlF (relative fluorescence units, RFU m$^{-2}$) and BB$_{700}$ (relative backscatter units, RBU m$^{-2}$) were computed as the depth-weighted sum of concentration per unit area, integrated from the top to bottom of a vertical zone. The saturation concentration of O$_2$ computed from temperature and salinity was subtracted from the observed O$_2$ to remove those physical impacts on gas concentration. For layers not being actively ventilated, this inverted apparent oxygen utilization (AOU) quantity, here referred to as "O$_2$ anomaly", resulted in positive dO$_2$/dt associated with "net autotrophy", and negative values with "net heterotrophy". A 3-week gap in NO$_3^-$ observations in May/June 2017 was filled by constructing a reference profile (NO$_3^-$ as a function of density) from the profiles on either side of the gap and projecting NO$_3^-$ onto the density profiles acquired by a replacement glider that was deployed in the same location between SUNA missions.

3.2 Net Community Production

Daily changes in O$_2$ and NO$_3^-$ inventories were used to independently estimate cumulative NCP in layers spanning the photic zone over the annual cycle in 2017. For O$_2$ in a layer from the sea surface to a depth, *h*, the mass balance equation is:

(1)
$$\int_{-h}^{0} \langle \frac{DO_2}{Dt} \rangle dz = NOP - ASE + F_{diff} - F_{ent} + \int_{-h}^{0} \langle u \frac{\partial O_2}{\partial x} + v \frac{\partial O_2}{\partial y} \rangle dz$$

That is, the bulk change in daily average O$_2$ in the layer – the term on the left side of [Eq. 1] – is the sum of net oxygen production (NOP) – a term that includes all phytoplankton-associated O$_2$ production and respiration terms regardless of nitrogen source and/or stoichiometry of the biomass produced – the vertical fluxes through the top and bottom of the layer (i.e. air-sea exchange (ASE), diffusive fluxes ($F_{diff}$) and entrainment flux ($F_{ent}$)), and the horizontal advective fluxes ($u \frac{\partial O_2}{\partial x} + v \frac{\partial O_2}{\partial y}$) in units of mol O$_2$ m$^{-2}$ d$^{-1}$. In this study, advective fluxes could not be accounted for because the sampling patterns that produced this dataset failed to provide a viable means of estimating them. Importantly, the thickness of the layer is allowed to fluctuate with time rather than assigning a fixed depth of integration (e.g. 150 m or the maximum depth of the winter ML are frequently used). This implementation enables the mass balance to be evaluated in sublayers by setting the limits of integration to depths $h_{top}$ and $h_{bot}$ determined, for example, from the time-varying boundaries of vertical zones. We applied this feature to estimate NOP for the photic layers (VZ$_0$, VZ$_1$ and VZ$_2$) and the subphotic layer (VZ$_3$) separately, and then as a whole, to quantify when, where, and how much net production occurred. Daily values of NOP were converted to NCP (g C m$^{-2}$) in accord with Redfield stoichiometry, C:O$_2$ = 106/138 and accumulated over an annual cycle to calculate ANCP. A lowpass filter (3$^{rd}$ order Butterworth with normalized cutoff frequency = 0.25) was applied to the time series of daily cumulative NCP and the resulting curves were used to evaluate the change in NCP for each season and layer, rounding the values to the nearest 5 g C m$^{-2}$.

The vertical flux components – ASE, F$_{diff}$ and F$_{ent}$ – were computed following established algorithms. Emerson and Bushinsky (2016) provided an outstanding comparison and explanation of gas exchange parameterizations which guided our implementations. When $h_{top}$ was the sea surface, ASE was calculated using the model by Liang et al. (2013) which incorporates mass transfer coefficients from the NOAA-COARE model of gas exchange and calculates bubble-induced gas fluxes from wind speed and a population distribution model. Wind speeds (10-m, hourly averaged) were derived from the ERA5 reanalysis products (Hersbach et al., 2018). When $h_{top}$ was not the sea surface, ASE was by definition 0, and $F_{diff}$, computed as the product of the mixing coefficient ($K_v$) and vertical $O_2$ gradient ($\partial O_2/\partial z$) was substituted for that term. At the base of $VZ_0$, $K_v$ was computed as a function of wind speed (Haskell et al., 2016); for layer boundaries below $VZ_0$, a canonical value of order $10^{-5}$ was applied (e.g. Lewis et al., 1986; Ledwell et al., 1993). An entrainment flux, $F_{ent}$, was calculated for $VZ_0$ and the underlying layer adjacent to it as follows. We set $MLD_1$ and $MLD_2$ to the depth of the ML at time = $T_1$ and $T_2$ respectively, and $H_1$ and $H_2$ to the thickness of the layer. Then $H_{ent} = |H_2 – H_1|$ is the thickness of the entrained layer, while $C_1$ and $C_{ent}$ are the mean concentrations of $O_2$ in $H_1$ and $H_{ent}$ respectively. The change in concentration in the layer at $T_2$, $dC_{ent}$, (due to entrainment of $H_{ent}$) and $F_{ent}$ are derived by Eqs. 2 and 3:

$$dC_{ent} = (C_{ent} – C_1) * H_{ent} / H_2 \qquad (2)$$

$$F_{ent} = dC_{ent} * H_{ent} / (T_2 – T_1) \qquad (3)$$

For a shoaling ML where the layer is detraining, $dC_{ent}$ is derived from $(C_1 – C_{ent})$. The sign of $F_{ent}$ for the layer beneath $VZ_0$ is the negative of $F_{ent}$ for $VZ_0$.

Analogous $NO_3^-$ terms were derived from the glider missions which carried a SUNA sensor and were used to produce estimates of net $NO_3^-$ utilization ($N_{util}$) rates for the photic zone in place of NOP. Again, no advective terms could be computed and were thus neglected. Here the integration limits were taken to be the sea surface down to the base of the euphotic zone – i.e. the base of $VZ_2$ during the *Spring*, *Stratified* and *Fall* seasons, or $VZ_0$ for the *Mixed* season. $F_{ent}$ and $F_{diff}$ were calculated as described above for NOP. Daily $N_{util}$ values were converted to NCP using Redfield C:N = 106/16, and their cumulative sums (same lowpass filter applied) were compared to NCP from NOP where the records overlapped in time. $NO_3$-based NCP will be underestimated by any discrepancies between our assumed conversion (C/N 106/16) from the actual C/N ratio in the produced biomass. C/N in the western North Atlantic (summary presented in Martiny et al. 2013) showed that the 25% and 75% data distribution limits were 5-7, with a median of ~5.8 and a mean of ~6.3. This is consistent with the general observation that the C/N ratio is much less variable than C/P and N/P ratios. Thus, in the way we have calculated the $NO_3$-based NCP, there is little impact of non-Redfieldian stoichiometry.

3.3 Discrete samples for validation and calibration of glider and FCM observations

Discrete samples for extracted chlorophyll, $O_2$, salinity, $NO_3^-$, and nano- and picophytoplankton were collected and analyzed using standard BATS methodologies (Lomas et al., 2013). In brief, to avoid compromising the samples by atmospheric gas exchange, dissolved $O_2$ samples were taken first, with 25% of the depths replicated including the surface, the deepest depth, and the oxygen minimum. Samples were analyzed using an automated Winkler titration method (Williams & Jenkinson, 1982) calibrated to a commercially available potassium iodate solution (OSIL, UK). Salinity samples were capped with a plastic insert to minimize evaporation and analyzed on a Guideline model 8400B Autosal Salinometer standardized against IAPSO Seawater Standards.

Samples for $NO_3^-/NO_2^-$ were gravity filtered and frozen (-20$^o$C) in HDPE bottles until analysis ashore (Dore et al. 1996). During every sample run, commercially available certified standards, OSIL and Wako Chemical, are analyzed to maintain data quality, as well as 'standard water' from 3000 m that serves as an internal standard. Bulk phytoplankton biomass (chlorophyll-a) was quantified by fluorometry (Yentsch and Menzel, 1963) on a calibrated Turner Designs Trilogy unit.

Samples for pico- and nanoplankton enumeration were collected, preserved (0.5% v/v paraformaldehyde) and flash-frozen in liquid nitrogen before being stored at -80$^°$C until analysis by flow cytometry). Upon analysis, populations of *Synechococcus* or *Prochlorococcus* and picoeukaryotes (<3 µm) and nanoeukaryotes (>3 µm) were enumerated as in Lomas et al. (2013). Phytoplankton cell abundance was converted to carbon per cell using a normalized cell size-carbon relationship and then to population biomass by multiplying by cell abundance (Casey et al., 2013).

The FCM biomass values obtained from discrete samples were interpolated to produce biomass profiles with 10-m vertical spacing and integrated from the sea surface to the base of $VZ_2$ to produce time-series estimates of population biomass within the photic zone over the years 2017 - 2018. Multiple linear regression using least squares was applied to determine scaling coefficients for the glider-based $BB_{700}$ and ChlF inventories (RBU and RFU integrated over the vertical photic zone) from total FCM phytoplankton biomass (eukaryotes + cyanobacteria) and eukaryote biomass, respectively.

## 4. Results and Discussion

4.1. ANCP from $O_2$ mass balance

The physical framework described in Section 2 provides a practical way to partition observational data into temporal and spatial groups that reflect fundamental processes underpinning the carbon cycle. It circumvents several complications of partitioning by fixed depths in the face of temporally varying water column structure, and refines the definitions of MLD, the base of the photic zone, and base of the maximum wintertime mixing based on density, ChlF and depth criteria. We have applied this framework to explore the timing and amplitude of NCP using glider observations acquired in 2017-2018.

Daily-average inventories and fluxes of $O_2$ were estimated for the photic and subphotic layers ($VZ_0$ – $VZ_3$) and their contributions to ANCP were evaluated. As noted in Section 3.2, these estimates do not account for advective fluxes which can be important (e.g., Alkire et al., 2014), particularly in the context of mesoscale eddies that characterize this study region. However, as 1-D models have been shown to capture most of the variability (e.g., Doney et al., 1996), this should not overly compromise the estimates. In our analysis, the effects of eddy advection manifested as short-term (10-14 day) bumps in cumulative NCP curves, primarily reflecting changes in thickness of subsurface layers and lateral $O_2$ gradients at eddy boundaries: e.g., the transit of a series of MWEs in *Fall* 2017 and a prominent CE at the end of the *Mixed* season in 2018 (Figs. 1 and 2). In the absence of realistic advection estimates, we have noted the timing of the eddies' presence and considered this factor in the evaluation of our results.

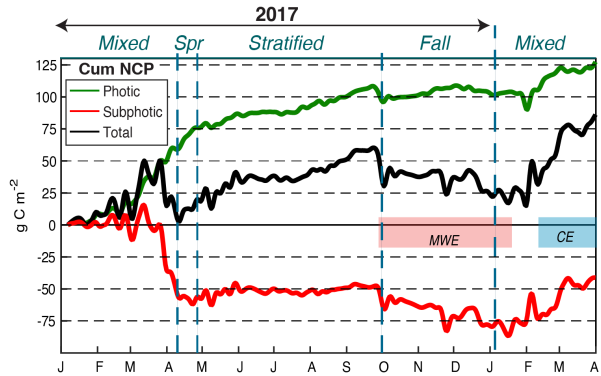

Figure 3. **Cumulative NCP from glider $O_2$ for the photic and subphotic layers in 2017 and early 2018**. Daily average NCP was computed from $dO_2/dt$ and converted to grams of carbon per unit area using Redfield $C:O_2$ = 106/138. The green curve reflects cumulative NCP for the photic layers (defined as $VZ_0+VZ_1+VZ_2$), the red curve is the subphotic layer $VZ_3$, the black curve is the sum of the photic and subphotic layers. Periods characterized by the presence of significant eddies (from Fig. 2) are denoted by horizontal colored bars.

For the full annual cycle (January 2017 to January 2018), $ANCP_{pho}$ – estimated from $O_2$ in the photic layers (i.e. $VZ_{0-2}$) – amounted to ~105 g C m$^{-2}$ y$^{-1}$ whereas $ANCP_{sub}$ – estimated for the subphotic layer ($VZ_3$) – exhibited a net decline of ~80 g C m$^{-2}$ (Fig. 3, green and red curves, respectively). Cumulative NCP broken down by layers and seasons are summarized in Table 3, and we point to the fact that these estimates have been rounded to the nearest 5 g C m$^{-2}$ yr$^{-1}$ reflecting our subjective assessment of their precision (i.e. on the order ±2.5 g C m$^{-2}$). Not surprisingly, the bulk of $ANCP_{pho}$ occurred at the tail of the *Mixed* season and into the *Spring* transition, accounting for approximately 55 and 20 g C m$^{-2}$ respectively, for a total of ~75 g C m$^{-2}$. In the subsequent *Stratified* period, an additional 30 g C m$^{-2}$ accumulated creating a peak at the end of September, while the *Fall* season was characterized by little further change. Mirroring $ANCP_{pho}$, the largest decrease in $ANCP_{sub}$ (~55 g C m$^{-2}$) occurred in the late *Mixed* season such that virtually all production above the MLD was quickly consumed beneath it resulting in zero cumulative NCP by the start of *Spring* (Fig. 3, black curve). The *Spring* and *Stratified* seasons, however, were characterized by little or no change in $ANCP_{sub}$ while the remaining decrease (~25 g C m$^{-2}$) occurred during the *Fall* season. Thus, for the combined photic and subphotic water column, $ANCP_{tot}$ amounted to approximately +25 g C m$^{-2}$ signifying that the region was net autotrophic for that year and implying an EP equivalent to that amount. The observed production and consumption exhibited distinct seasonality: net zero NCP (within rounding errors) accumulated over the *Mixed* season, positive NCP accumulated in the *Spring* and *Stratified* seasons, while the system reverted to net heterotrophy in *Fall* driven by $O_2$ consumption in the $VZ_3$ layer between the base of the photic zone and density of the deepest winter mixing. Moreover, the amplitude of $ANCP_{tot}$ was consistent with previously cited direct estimates of EP (from POC + Fz + DOM), i.e., 22.1±8.8 g C m$^{-2}$ y$^{-1}$.

Table 3. Cumulative NCP by layers and seasons (rounded to nearest 5 g C m$^{-2}$)

| Cum NCP Season | Photic (g C m$^{-2}$) | Subphotic | Photic + Subphotic |
|---|---|---|---|
| Mixed | +55 | -55 | 0 |
| Spring | +20 | 0 | +20 |
| Stratified | +30 | 0 | +30 |
| Fall | 0 | -25 | -25 |
| Annual | +105 | -80 | +25 |

How has the lack of advective fluxes affected this analysis of ANCP? A series of MWEs transiting the sampling site during *Fall* (colored bar, Fig. 3) corresponded to brief declines in both $NCP_{pho}$ and $NCP_{sub}$ which reflect the simultaneous uplift of the $VZ_3$ upper boundary and deepening of its lower boundary. Each of those short-term (12-14 day) decreases (in the range 5-15 g C m$^{-2}$) rebounded somewhat, although not completely, after the eddy front passed by the sampling site. This makes sense, because lateral gradients of $O_2$ that modulate the magnitude of advective fluxes are more pronounced across the eddy edges compared to the eddy interior where they are generally small. The largest impact, at the beginning of October, was associated with a net decline in $NCP_{tot}$ of ~15 g C m$^{-2}$ over 14 days (~1.1 g C m$^{-2}$ d$^{-1}$), a rate that is ~4 times higher than the linear rate of decline over the remainder of the *Fall* season (~0.25 g C m$^{-2}$ d$^{-1}$). Applying this latter rate over the entire *Fall* period (90 days) elevates the estimate of $ANCP_{tot}$ by ~12 g C m$^{-2}$ yielding a range of values: 25-37 g C m$^{-2}$ y$^{-1}$. Thus, these fluctuations represent a source of noise of order 10-15 g C m$^{-2}$ that must be included in assessing uncertainties; but they do not significantly alter our interpretation of cumulative NCP trends on seasonal timescales.

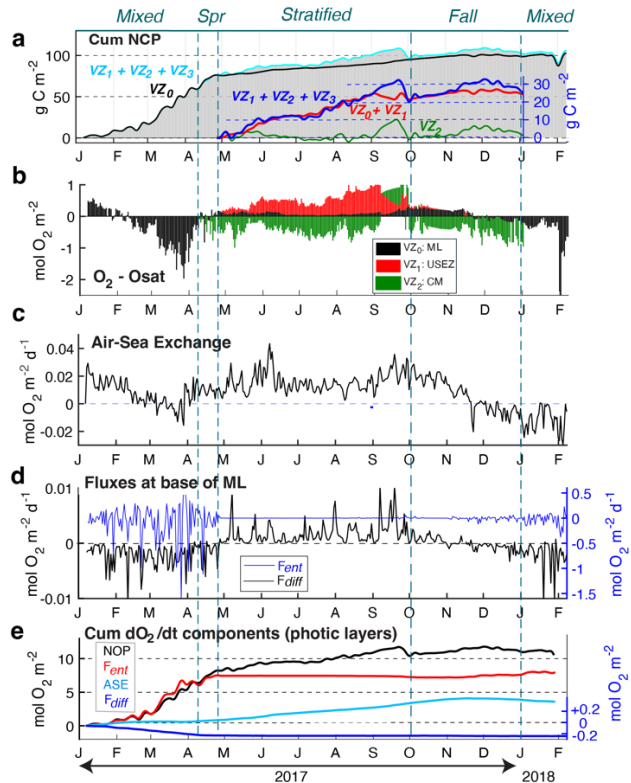

*Figure 4.* **Cumulative NCP, $O_2$ inventories and fluxes in vertical zones spanning the photic layers.** **(a)** The gray shaded area and cyan line reflect the sum of $VZ_0$, $VZ_1$ and $VZ_2$ over 13 consecutive months, with the contribution from $VZ_0$ depicted by the black line (scale is the vertical axis left, 0:120 g C m$^{-2}$). For the *Stratified* and *Fall* seasons, $NCP_{pho}$ is divided into upper and lower layers (scale is the blue vertical axis on the right, -2:30 g C m$^{-2}$): $VZ_0 + VZ_1$ (red), $VZ_2$ (green), and their sum (blue). **(b)** $O_2$ inventory (concentration integrated over thickness of layer) with the saturation concentration removed. Color differentiates layers: $VZ_0$ (black, Mixed Layer (ML)), $VZ_1$ (red, upper stratified euphotic zone (USEZ)), $VZ_2$ (green, chlorophyll maximum layer (CM)). Positive values signify supersaturation and negative values are undersaturated. **(c)** Air-sea exchanges of $O_2$ (computed following Emerson and Bushinsky, 2016). **(d)** Diffusive and entrainment fluxes through base of $VZ_0$ – note separate Y-axes and scales. **(e)** Cumulative NOP (black line) for the photic layers (mol $O_2$ m$^{-2}$), with relative contributions to $dO_2/dt$ from vertical fluxes through the top and base of photic zone: $F_{ent}$ (red), ASE (cyan) and $F_{diff}$ (blue). Left axis (black) is for NOP, $F_{ent}$ and ASE (scale is 0-12 mol $O_2$ m$^{-2}$); right axis (blue) is for $F_{diff}$ (scale is -0.2 - +0.2 mol $O_2$ m$^{-2}$).

Next we consider the partitioning of $NCP_{pho}$ over the course of the *Stratified* season. Inventories of $O_2$ anomaly highlight a sharp contrast between supersaturated and undersaturated concentrations in $VZ_1$ and $VZ_2$, respectively, with much smaller amounts in $VZ_0$ (Fig. 4b). Approximately 80% of the ~30 g C m$^{-2}$ of $NCP_{pho}$ was produced in the shallower layers, $VZ_0 + VZ_1$, while $VZ_2$ exhibited negligible values until September, when $VZ_1$ was mixed away and conveyed its supersaturated waters to adjacent layers (Fig. 4a, right axis). Moreover, changes in $NCP_{pho}$ were small over the *Fall* season (< 5 g C m$^{-2}$) and accumulated mainly in the deeper layer, $VZ_2$, opposite to the situation in previous months. It is worth noting that NCP was negligible in the shallow photic layers ($VZ_0 + VZ_1$) during *Fall* despite the deepening ML, because stratification at the base of $VZ_2$ inhibited nutrient fluxes to the photic zone from below and phytoplankton biomass declined to minimum values (Section 4.4).

Daily average vertical fluxes through the top and bottom of $VZ_0$ and their relative contributions to cumulative $dO_2/dt$ emphasized the importance of entrainment, as well as its episodic nature, during the *Mixed* and early *Spring* seasons (Fig. 4). The amplitude of $F_{ent}$ was 1-2 orders of magnitude greater than $F_{diff}$ and closely tracked the accumulation of NOP from 0 to 7 mol $O_2$ $m^{-2}$ in $VZ_0$, while $F_{diff}$ was downward and accounted for approximately -0.2 mol $O_2$ $m^{-2}$. Daily ASE was generally positive in sign (out of the ocean) except at the end of the *Mixed* season and the last half of *Fall* at which time cumulative ASE reached a peak value of ~+4.8 mol $O_2$ $m^{-2}$ (Fig. 4c, e). In contrast, air-to-sea $O_2$ fluxes were observed in March, coinciding with an abrupt deepening of the ML from 200 to 300 m (Fig. 1a) associated with a series of strong wind events and convective mixing, whereas cooling surface temperatures gradually reversed the direction of the flux in November and December.

In the underlying $VZ_3$ layer, $NCP_{sub}$ exhibited significant declines indicating $O_2$ consumption through respiration and remineralization processes in the late *Mixed* season and throughout *Fall*, but little net change (0 - 5 g C $m^{-2}$) during the *Spring* and *Stratified* seasons (Fig. 3). However, partitioning the layer into upper and lower sublayers divided by the isopycnal $\sigma_0 = 26.20$ kg $m^{-3}$ revealed a curious sign reversal: NCP in the shallower sublayer exhibited an *increase* equivalent to ~15 g C $m^{-2}$ between the start of *Spring* and the end of May, while the deeper sublayer simultaneously declined by ~10 g C $m^{-2}$ (Fig. 5a). Observations of this phenomenon in subsequent years (2021, 2023, 2024) were linked to diel vertical migrations of large salp blooms whose nightly excursions from the twilight layers up to graze on phytoplankton in the CM layer generated turbulence in the WMW layer. Those datasets will be published separately, but it is worth noting that a key signature in those years included enhanced turbulent dissipation of energy (measured by gliders equipped with microturbulence sensors) occurring nightly between the photic zone and 400 meters depth, and diapycnal mixing of properties (e.g. $O_2$ and salinity) across the photic / subphotic layer interface. In 2017, similar perturbations also suggested vertical mixing of $O_2$-rich waters from the photic layers with $O_2$-depleted waters from below, rather than changes in subphotic NCP rates. Moreover, salp abundances – determined from a combination of metabarcoding and imaging techniques applied to zooplankton tow samples – were found to be extremely high at BATS (within 10 km of the glider) in April of that year (Perhirin et al., 2024).

Diapycnal exchanges were clearly evident in April/May 2017 after convective mixing had ended – i.e. at the start of *Spring* when the base of $VZ_2$ and isopycnals in the upper and lower $VZ_3$ sublayers (26.20 and 26.30, respectively) were vertically separated in depth (yellow shading, Fig 5). Salinity, which is generally conserved along isopycnals, fluctuated on 26.20 between the values at the base of $VZ_2$ and the deeper isopycnal at diel frequencies (Fig. 5c). $O_2$ and $NO_3^-$ anomalies exhibited analogous variability on the isopycnals suggesting physical redistribution between the properties in $VZ_2$ and $VZ_3$ as opposed to biological production or uptake in the subphotic layer (Fig. 5d,e). This phenomenon was associated with a rise in $NCP_{tot}$ equivalent to ~30 g C $m^{-2}$ from the start of *Spring* to the end of May – i.e. approximately 50% of the total rise in $NCP_{tot}$ over the entire *Stratified* season (Fig. 3). If indeed these were associated with a salp bloom, we speculate that these increases may have reflected enhancement of both EP (fecal pellet production) and NP (diapycnal $NO_3^-$ flux). This leads us to further propose that the inability to predict the strength of the export pathway (NCP:NPP ratios) based on atmospheric forcing factors alone (Brix et al., 2006) may have been linked to the existence and impacts of biologically-mediated delivery of $NO_3^-$ and EP in the weeks after deep convective activity had ceased.

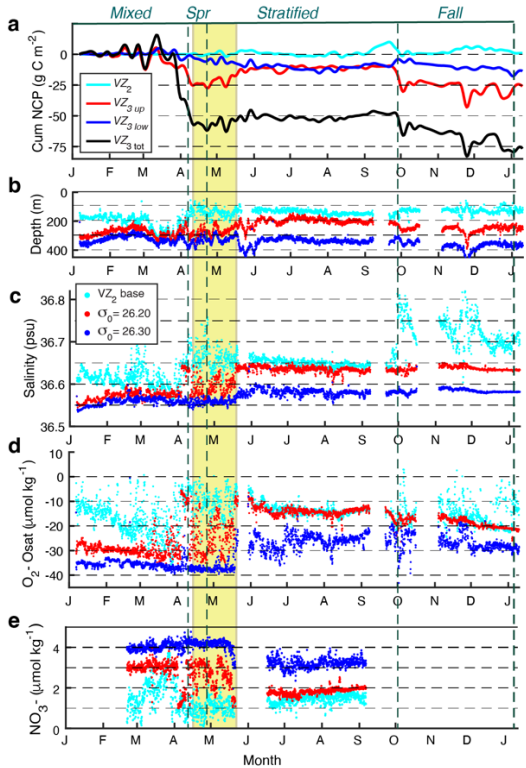

*Figure 5.* **Cumulative NCP and properties on density levels in the subphotic layer, $VZ_3$.** (**a**) Cumulative NCP curves computed from $dO_2/dt$ in sublayers (legend in lower left corner): $VZ_2$ (cyan), $VZ_3\,up$ (i.e. base of $VZ_2$ to $\sigma_0 = 26.20$, red), $VZ_3\,low$ ($\sigma_0 = 26.20$ to base of $VZ_3$, blue) and $VZ_3\,tot$ (total for the $VZ_3$ layer, black). (**b-e**) Point observations of depth, salinity, $O_2$ anomaly and $NO_3^-$ at three levels (see legend in panel **c**): base of the photic zone (cyan) and on the isopycnals 26.20 (red) and 26.30 (blue) kg m$^{-3}$. Brief gaps in the observations reflect sensor failures which were not interpolated over. Yellow shading highlights a period characterized by enhanced vertical exchanges between the photic and subphotic layers for several weeks *after* winter convection had ceased. The mixing is attributed to large blooms of salps whose nightly vertical migration patterns from the twilight layers up to graze on phytoplankton in the DCM layer generated turbulence in the WMW layer. A "rise" in NCP in the upper portion of $VZ_3$ (panel **a**) during the first 4 weeks of the *Stratified* season reflects the physical mixing of high $O_2$ concentrations from above with waters depleted in $O_2$ from below (panel **d**), rather than $O_2$ production in the subphotic layer. These diapycnal exchanges are echoed in salinity and $NO_3^-$ concentrations (panels **c** and **e**, respectively).

4.2 Cumulative NCP from $NO_3^-$ mass balance

The $NO_3^-$ time series is less complete than $O_2$ reflecting a period of repairs to the SUNA sensor from September to December, but over the course of its missions, the sensor measured continuously at a vertical resolution of ~4 m, in contrast to discrete sampling associated with the BATS program which measures at a vertical resolution of ~20 m. A short gap between SUNA deployments (21 May – 15 June) was filled by interpolation of a reference density:$NO_3^-$ curve derived from the daily-averaged profiles acquired on 20 May and 16 June onto the density profiles measured by another glider during the gap. $NO_3^-$ inventories (mmol N m$^{-2}$) and vertical fluxes were computed for the photic layers, and the mass balance equation for $dN/dt$ was used to evaluate cumulative $NCP_{pho}$ for the January to September time frame (Fig. 6).

Daily average inventories highlighted the episodic nature of $NO_3^-$ delivery and convective mixing from January to early April, and the much lower levels of nutrients present in the photic layers after the system stratified (Fig. 6b). $F_{ent}$ accounted for nearly all $NO_3^-$ flux (~730 out of 760 mmol N m$^{-2}$) but plateaued in mid-April, while $F_{diff}$ contributed just ~20 mmol N m$^{-2}$ over the same time period and delivered an additional ~6-7 mmol N m$^{-2}$ over the course of the *Stratified* season (Fig. 6c). There was no evidence of diel (nighttime vs daytime) signals in $NO_3^-$ inventories in those months. Moreover, we expect that a similar amplitude of $NO_3^-$ flux continued through *Fall* because MLD did not penetrate below the base of $VZ_2$ precluding entrainment of nutrient-rich waters into the euphotic zone. Thus we can infer another ~3 mmol N contribution in the later part of the year for an annual total of ~760 mmol N m$^{-2}$ y$^{-1}$.

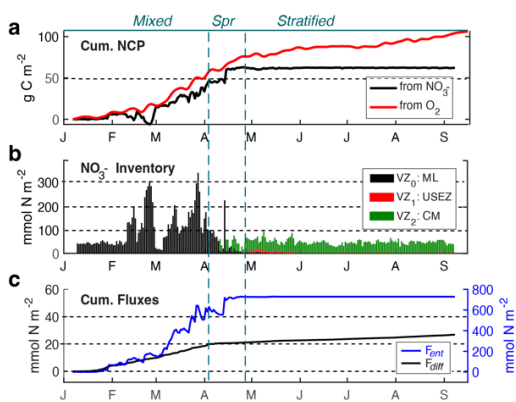

*Figure 6.* **NCP, $NO_3^-$ and fluxes through base of the euphotic zone in 2017. (a)** Cumulative NCP computed from $NO_3^-$ profiles (converted to carbon equivalents using Redfield C:N = 106/16) and compared to NCP from $O_2$ (see Fig. 3) for an overlapping segment of observations from January to September 2017. **(b)** $NO_3^-$ inventories (concentration integrated over thickness of layer) for individual vertical zones spanning the photic layers as in Fig. 4. **(c)** Cumulative fluxes of $NO_3^-$ through the base of $VZ_2$ (or $VZ_0$ in the *Mixed* season): diffusive flux (black line, left axis) and entrainment flux (blue line, right axis), note difference in scales.

Cumulative $NCP_{pho}$ derived independently from $NO_3^-$ and $O_2$ (Fig 5a) closely tracked each other in rate and amplitude through the *Mixed* and early *Spring* seasons accounting for ~60 g C m$^{-2}$ or slightly more than half of the $ANCP_{pho}$. The continued rise in $NCP_{pho}$ during the late *Spring* and *Stratified* months, however, had no counterpart in NCP computed from $NO_3^-$. Moreover, the small amounts of $NO_3^-$ measured in the photic layers were mainly restricted to $VZ_2$, whereas the bulk of $NCP_{pho}$ accumulated in $VZ_0 + VZ_1$. This implies biological, rather than dynamical, controls on the vertical distributions of $NCP_{pho}$ during the *Stratified* and *Fall* seasons.

4.3 Comparisons to previous estimates

With daily accounting of upper ocean properties and vertical structure, these glider observations provide insights that are both novel and complementary to previous efforts to quantify carbon exports. To facilitate comparisons to earlier studies, we convert the units in our NCP estimates from mass to molar equivalents. Compilations of geochemical estimates of NCP at BATS present a range of values: e.g. from 2-5 mol C m$^{-2}$ y$^{-1}$ including averages of 3.8±1.2 mol C m$^{-2}$ y$^{-1}$ (Emerson, 2014) and 3.0±1.0 mol C m$^{-2}$ y$^{-1}$ (Fawcett et al. 2018). The individual estimates employed a variety of techniques, temporal and vertical sampling resolution, and integration over assorted depth ranges. And while differences from direct measures of carbon export (i.e. 1.8±0.5 mol C m$^{-2}$ y$^{-1}$) have raised concerns about inaccuracies in the latter (Lomas et al., 2013; Emerson, 2014; Fawcett et al. 2018), the implied disconnect between ANCP and EP may be less pronounced and arise from other factors than previously suggested.

For the annual cycle in 2017, glider-based estimates of ANCP in the combined photic and subphotic layers (including uncertainties due to lack of advective fluxes) summed to 2.5±0.5 mol C m$^{-2}$ y$^{-1}$, an amplitude that is (within error bars) consistent with direct estimates of EP. Independent measures of $NCP_{pho}$ from $O_2$ and $NO_3^-$ tracked each other during the *Mixed* season when episodic fluxes of nutrients fueled production in $VZ_0$ and were offset by consumption in the underlying layer ($VZ_3$), leading to net zero EP by the start of *Spring*. Daily-average $NO_3^-$ inventories in the photic layers dropped precipitously following the onset of stratification and remained quite low for the remainder of the year summing to ~10 mmol m$^{-2}$. Despite the diminished supply of $NO_3^-$, $NCP_{pho}$ continued to accumulate during the *Spring* and *Stratified* seasons (principally in $VZ_0 + VZ_1$), but exhibited little change during *Fall*, whereas $NCP_{sub}$ exhibited little change during the *Stratified* season but declined during *Fall*. These patterns are consistent with findings by Fawcett et al. (2018), who attributed unexpectedly high $O_2$ production and DIC drawdown, in the absence of a physical nutrient supply, to production of organic matter with much lower N and P content than is typical of Redfield dynamics.

Fawcett et al., 2018 suggested a vertical separation of $O_2$ production and consumption, mediated by a flow of carbon into nutrient-depleted gel-like organic matter (GLOM) that slowly sinks and is respired in deeper layers by heterotrophic bacteria. They proposed that under conditions of nutrient limitation in the summer and fall at BATS, $O_2$ production occurs near the surface of the stratified euphotic zone (similar to our $VZ_0+VZ_1$), and remineralization of GLOM occurs between the summer euphotic zone depth and wintertime maximum MLD (synonymous with our $VZ_3$). They argued that the production of GLOM and its export beneath the euphotic zone would account for the high rates of ANCP derived from geochemical estimates (i.e. 3–3.8 mol C m$^{-2}$ y$^{-1}$). Furthermore, because GLOM requires a minimal supply of nutrients, their hypothesis addresses one of the essential problems in the BATS carbon budget: high rates of NCP in the absence of an observable source of nutrients during the stratified months. From 4+ years of profiling float data they compared the rates of $O_2$ decreases to nitrate increases at 130m (analogous to our $VZ_3$) and inferred greatly reduced $NO_3^-$ production rates i.e. <10% of the amount expected from Redfield stoichiometry (Figure 3b and Table S2 in Fawcett et al. 2018). They attributed this to decomposition of low nutrient DOM at "shallow" depths below the euphotic zone (in our $VZ_3$). On a deeper horizon – the 26.4 isopycnal which lies below the WMW layer in our $VZ_4$ – they estimated that $NO_3^-$ production rates were closer to 75% of Redfield stoichiometry (their Figure 3d), thus implying that a recycling of GLOM occurs between the summer euphotic zone and maximum winter ML depths (their Figure 1). This observation is consistent with incubation experiments which demonstrate that subphotic zone bacteria are able to degrade certain forms of DOM produced in surface waters more efficiently than surface zone bacteria (Carlson et al. 2002). This GLOM mechanism could contribute to discrepancies between NCP estimates and measured EP observations, as the sediment traps do not capture GLOM-rich water. Indeed any GLOM-rich water in the sediment trap collection tubes is syphoned off so that only particulate matter is quantified in export flux (BATS methods manual). The microbial carbon pump concept is a similar mechanism for funneling newly fixed carbon into nutrient depleted organic matter pool (e.g. Jiao et al., 2024) but likely acts on different timescales of remineralization relative to the GLOM-model.

The glider data support this picture and provide a magnitude for the shallow remineralization loop equivalent to ~25 g C m$^{-2}$ y$^{-1}$ reflecting production of organic matter that may be substantially non-Redfieldian (GLOM) in the upper photic layers ($VZ_0 + VZ_1$). (This is in addition to the ~55 g C m$^{-2}$ that was produced and remineralized during the *Mixed* season in 2017.) Production primarily occurred during the *Stratified* months while remineralization occurred mainly in the *Fall* season in the subphotic layer ($V_3$), implying a time lag for the downward transport of GLOM. This form of carbon export and recycling above the depth of wintertime mixing does little to reconcile ANCP with EP and its role in the biological carbon pump as the $CO_2$ from subsurface remineralization could be vented back to the atmosphere in winter rather than stored in the deep ocean. However, the glider data indicate an additional ~25 g C m$^{-2}$ y$^{-1}$ was produced, but not consumed, in the combined photic + subphotic layers during the *Stratified* months and therefore exported to depths beneath $VZ_3$. This number more closely approximates the magnitude of direct estimates of EP, but also the amount of NCP that accumulated in conjunction with diapycnal mixing of $O_2$ and salinity across the photic / subphotic layer interface (inferred to be biologically generated) in April and May. Thus, the overall picture strongly resembles the dual system proposed by Brix et al. (2006) of an export pathway reflecting new production in addition to a regeneration loop fueled by recycled nutrients. Key insights include our speculation regarding the role of biological communities in delivering the source of $NO_3^-$ and EP in the *Spring* and early *Stratified* seasons (beyond the NP fueled by winter convective mixing), while NCP in the second half of the year occurs with minimal nutrient input and low EP in the portion of the photic zone dominated by cyanobacteria communities.

4.4. Relationships to phytoplankton community structure

Elemental stoichiometry of phytoplankton has received considerable attention and it is now well accepted that macronutrient stoichiometric ratios can vary dramatically between taxa (e.g., Geider and LaRoche, 2002; Quigg et al., 2003, Ho et al., 2003), and within taxa, as a function of growth conditions (e.g., Leonardos and Geider 2004, Whitney and Lomas 2016, Garcia et al 2018). Relevant to the nutrient-poor Sargasso Sea, cyanobacteria commonly display C:P and N:P ratios several-fold greater than the Redfield ratio (Bertilsson et al. 2003, Baer et al. 2017). Spatial variability in phytoplankton populations has been postulated to constitute an important driver of observed patterns in bulk particulate stoichiometry (Martiny et al. 2013; Sharoni and Halevy, 2020; Lomas et al. 2021), and global patterns in primary productivity and connectivity between trophic levels (Moreno and Martiny 2018, Kwiatkowski et al. 2018).

To what extent does vertical ecosystem structure in phytoplankton communities imprint on the biogeochemistry in this region? Profiles of phytoplankton biomass derived from FCM of discrete water samples in 2017 and 2018 showed expected seasonal variations in eukaryote vs. cyanobacteria populations over the annual cycle, Fig. 7a (Casey et al. 2013). For most of the year cyanobacteria groups constituted the bulk of biomass – i.e *Synecoccochus* dominated the early *Mixed* season, while *Prochlorococcus* were predominant during the *Stratified* and *Fall* seasons. Biomass of the eukaryote groups rose in the *Mixed* and *Spring* seasons in response to episodes of $NO_3^-$ entrainment. The flow cytometric method employed in this study does not allow for the effective identification of diatoms that might also accumulate during the *Mixed* season. Due to the lack of more direct measurements, the contributions of large phytoplankton, frequently diatoms, was assessed by the difference in total phytoplankton carbon and flow cytometric derived phytoplankton carbon (see Figure 4c in Lomas et al. 2022) and was found to be rare during the period of this study. For this reason we believe that growth of diatoms made only a minor contribution to biomass and subsequent impacts on ANCP.

The FCM estimates of total and eukaryote biomass were applied to scale the glider's optical sensors producing a ratio (ChlF:$BB_{700}$) which we utilize as an index for relative abundances of eukaryotes and cyanobacteria groups (Fig. 7b, c). For the sum of the photic layers $VZ_{0-2}$, the index indicates that eukaryote biomass ranged from 20 – 50% of total biomass over the annual cycle in 2017 – highest and quite variable in the *Mixed* and *Spring* seasons, near 40% for most of the *Stratified* months, then declining to ~20% in *Fall*. Partitioned by individual layers, however, the index highlighted a striking vertical zonation during the *Stratified* months: the base of the photic zone, $VZ_2$ where little net $O_2$ production occurred, was characterized by eukaryote abundances ranging 50-80%, while in the shallower layers, $VZ_0$ and $VZ_1$, where most of the net $O_2$ production occurred, *Prochlorococcus* represented ~85% of phytoplankton biomass (Fig. 7c). It suggests that vertical gradients in $NO_3^-$ availability strongly influence phytoplankton community structure, which in turn leads to the vertical disconnect between $O_2$-based and $NO_3^-$-based NCP measured by the gliders.

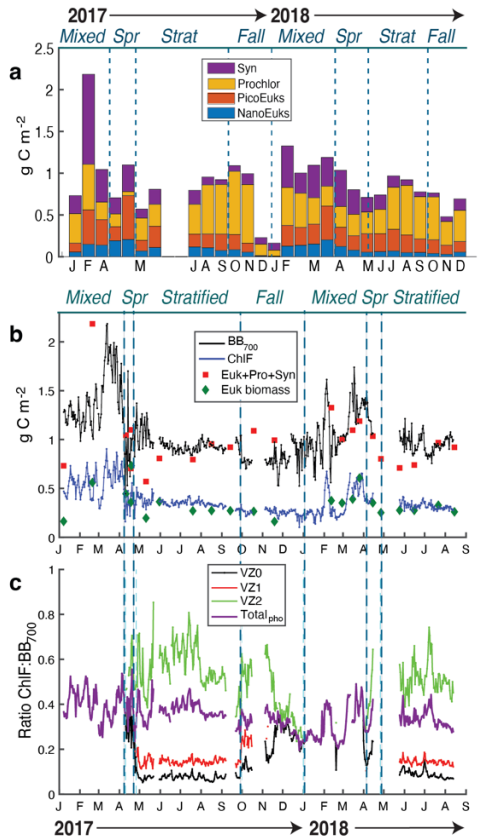

*Figure 7.* **Phytoplankton biomass measured by FCM and optical ChlF and backscatter measured by gliders.** (**a**) Biomass (g C m$^{-2}$) from FCM in photic zone by cruise (month) over a 2-year span differentiated by phytoplankton groups: *Synechococcus*, *Prochlorococcus*, nano-eukaryotes and pico-eukaryotes. (**b**) Glider $BB_{700}$ and ChlF optical properties scaled to FCM biomass ($BB_{700}$ to total euphotic phytoplankton biomass (eukaryotes + cyanobacteria); ChlF to eukaryote biomass). (**c**) Timeseries of the ChlF: $BB_{700}$ ratio differentiated by vertical zones ($VZ_0$, $VZ_1$, $VZ_2$ and combined total photic zone) as an indicator of seasonal and depth varying phytoplankton community structure.

Changes in phytoplankton biomass during the *Stratified* season (Fig. 7a) indicated net growth of phytoplankton equivalent to ~0.16 g C m$^{-2}$ in 2017, and ~0.1 g C m$^{-2}$ in 2018, which translates to ~0.2 g $O_2$ m$^{-2}$ production and ~0.13 g $O_2$ m$^{-2}$ of net production respectively (Richardson and Bendtsen, 2017). Thus, it appears that ~50% of cumulative NCP increase over the *Stratified* period was incurred from August to October and corresponded to net growth in phytoplankton dominated by *Prochlorococcus* populations. Based on averaged C:N ratios (Baer et al. 2017) only 0.02 g N m$^{-2}$ of N in 2017 and 0.01 g N m$^{-2}$ in 2018 were needed to support this net biological growth. Estimates of $F_{diff}$ for the same period are substantially higher (e.g. 0.06 g N m$^{-2}$ in 2017) suggesting that nutrient availability was sufficient to support the observed growth. The excess further implies a potential role for other vertical processes – e.g. trophic export and grazing – to influence NCP and EP in the later part of the year.

Although the degree to which small phytoplankton such as cyanobacteria contribute to the vertical transport of particulate matter remains under debate (e.g., Lomas and Moran, 2011; Letscher et al., 2023), it is generally accepted that they do contribute via different trophic pathways, e.g., salps (Stone and Steinberg, 2016). Salps are known to graze effectively on phytoplankton as small as cyanobacteria thus providing a mechanism to imprint the non-Redfield stoichiometry of phytoplankton onto the broader biogeochemical signal. Additionally, vertical profiles of microzooplankton grazing rates suggest that their grazing exceeds growth below the euphotic zone and is generally more balanced in summer than in winter when growth exceeds grazing (De Martini , 2016). Indeed, vertical gradients of phytoplankton growth patterns, non-Redfield stoichiometry, and grazing would all tend to reinforce the vertical disconnect between $O_2$-based and $NO_3^-$-based NCP and enable the Sargasso Sea ecosystem to continually produce and export C despite its nutrient impoverished state (Lomas et al. 2022). To the extent that cyanobacteria contribute to carbon export – directly through gravitational sinking or indirectly through trophic interactions – their role in particulate nutrient biogeochemistry would be synergistic with the dissolved organic nutrient biogeochemistry proposed by Fawcett et al. (2018). Moreover, it is inferred from Fawcett et al. 2018 that a highly active 'remineralization zone' exists below the euphotic zone, while Lomas et al. (2022) estimated that ~80% of the increase in particulate stoichiometric ratios in shallow sediment traps was due to rapid remineralization rather than changes in stoichiometry of source material during its vertical transit. Thus, the sum of biological and glider data presented here, in the context of other biogeochemical data at BATS, supports and expands Fawcett and colleagues' hypothesis for a role of non-Redfield biogeochemistry in maintaining and amplifying vertical disconnects in NCP.

5. Conclusions

For the annual cycle in 2017, we have quantitatively assessed ANCP, nutrient fluxes, and EP from mass balances of $O_2$ and $NO_3^-$ observations acquired with underwater autonomous gliders in coordination with the BATS program. Partitioning the data into a framework of seasons and vertical layers that reflect dynamical, physical, and biogeochemical boundaries in the Sargasso Sea, this analysis has produced estimates of the amplitude, timing, and vertical distribution of net $O_2$ production and consumption in the upper ocean, the rates and cumulative amounts of $NO_3^-$ delivery, the amount of carbon exported to the ocean interior beneath the depth of winter mixing, and the amounts recycled above it. Daily average inventories of physical and biogeochemical properties enabled us to evaluate processes over the annual cycle at greatly enhanced resolution compared to monthly ship-based observations. These results affirm and weave together several previously proposed ideas about the nature of the system, and reduce uncertainties attributed to sampling frequencies.

Analogous to the Brix et al. (2006) hypothesis, an "export pathway" fueled by mechanical delivery of $NO_3^-$ to the photic layers dominated the first half of the year and was followed by a "regeneration loop" in the subsequent *Stratified* and *Fall* seasons when the photic layers were nutrient deficient. Production and remineralization associated with the latter were separated in both space and time supporting the Fawcett et al. (2018) hypothesis for non-Redfieldian generation of nutrient-poor organic matter in the upper photic layers that sinks and is remineralized below the base of the photic zone. At those depths, subphotic microbial communities are able to degrade forms of "recalcitrant" organic matter more efficiently than surface communities (Carlson et al. 2002). Vertical zonation and seasonality in phytoplankton community structure – i.e. biomass of eukaryotes vs. cyanobacteria populations derived from FCM and used to scale the gliders' optical sensors – aligned with the patterns of $NO_3^-$ availability, $O_2$ production and consumption providing complementary evidence in support of this overall picture.

Mechanistically diverse factors governed the strong seasonality observed in NCP, NP and EP. Atmospheric forcing, intermittent convective mixing, and entrainment of nutrient-rich waters into the ML led to production of ~55 g C m$^{-2}$ that was quickly consumed during the *Mixed* season resulting in zero cumulative NCP and EP by the onset of stratification in April. The *Spring* and early *Stratified* months saw a rise in NCP which we speculate may have been associated with vertically migrating organisms, leading to observed diapycnal exchanges of water properties between the photic and subphotic layers. This process in 2017 contributed NCP equivalent to 25-30 g C m$^{-2}$ that was exported to depths beneath the WMW layer. The specific causes, frequency and spatial distribution of this phenomenon merit more explicit exploration. A third rise in NCP, also of order 25-30 g C m$^{-2}$, occurred from August to October, despite a much-reduced supply of $NO_3^-$ to the photic layers. This corresponded to accumulation of supersaturated $O_2$ concentrations in the shallow photic layers ($VZ_0 + VZ_1$) dominated by *Prochlorococcus* populations, in stark contrast to the base of the photic zone ($VZ_2$), closer to the slow, but steady, diffusive flux of $NO_3^-$ where eukaryotes were relatively more abundant but little NCP occurred. The magnitude of NCP that accumulated in the photic layers was approximately equivalent to the decline observed in the subphotic layer over the *Fall* season, signifying the potential for recycling the material within reach of winter mixing during the following *Mixed* season.

While this dataset reflects just a single year, over the full annual cycle, NCP was equivalent to ~105 g C m$^{-2}$. Of that amount, ~80 g C m$^{-2}$ was remineralized above the base of $VZ_3$ while ~25 g C m$^{-2}$ was exported to deeper layers. This yields an export ratio of ~25% -- higher than the canonical ~10% value for the global ocean (Eppley and Peterson 1979; Lomas et al. 2013) and closer to the geochemical estimates cited in the introduction (Lipshultz et al., 2002; Brix et al., 2006; Emerson, 2014; Fawcett et al., 2018). These findings suggest that oligotrophic ocean gyres, which have a vertically structured ecosystem, may be more important for carbon sequestration than generally believed. Maintaining continuous glider observations could provide valuable insights regarding interannual physical variability as well as biological/ecosystem variability, e.g., exploring the role of salps in the local carbon cycle (nutrient delivery and EP). Moreover, information regarding the resilience of oligotrophic gyres in the context of hypothesized reductions in carbon sequestration is increasingly germane to anticipating their response to globally changing climate conditions (e.g. Bopp et al., 2013).

**Data statement**
The post-processed glider data are archived as netcdf files (individual profiles sorted by time) at the IOOS U.S. Glider DAC available via their data portal at https://www.ncei.noaa.gov/access/integrated-ocean-observing-system/. The observational data pertinent to this paper are also distributed online via a project-specific folder maintained at Arizona State University:
https://www.dropbox.com/scl/fo/jjp8248xeioy71c4kldub/ALKjmb7Ri-KnMks2HWvKjs4?rlkey=o5memst972gyrn0mvxuvlqiek&dl=0.
These include the glider data stored by mission in netcdf files of trajectory profile data and the daily-averaged profile data stored in a Matlab format. Discrete water sample data are stored in .xlsx and .mat files. Wind data (from ERA5) are stored in a Matlab format.

**Author Contributions**
Glider operations, data processing, data curation and analyses were performed by RGC. MRS contributed to glider data analysis as part of a student internship at BIOS in 2019. CTD and water sample data were collected at sea by the BATS team in 2017/2018. MWL performed the FCM and analyses. All authors contributed to the completion of the manuscript.

# Appendix A.  Supplementary data

**Acknowledgements**

This work was supported by US NSF through OCE-1851224 to RGC and DG and OCE-1850723 to MWL. MRS received support from the JHU student internship program.  We also gratefully acknowledge support for the BIOS glider program from the G. Unger Vetlesen Foundation, Grayce B. Kerr Fund and Simons Foundation International.  We are indebted to the BATS team who assisted with glider deployments, recoveries and provided CTD and water sample data used to calibrate the glider sensors. We also thank the officers and crew of *R/V Atlantic Explorer* and the BIOS Small Boats Team for their assistance with glider field operations.

Supplement

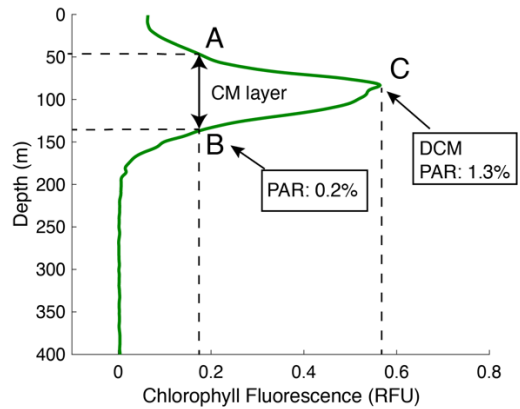

**Figure S1. Definition of CM layer from a chlorophyll profile.** The upper and lower bounds of the layer (**A** and **B** respectively), correspond to the depths where chlorophyll is 0.35 of the maximum value at **C**, the DCM. Mean PAR/$PAR_0$ percentages at **B** and **C** were estimated from multiple glider missions carrying a PAR sensor (but not limited to the 2017/2018 observations).

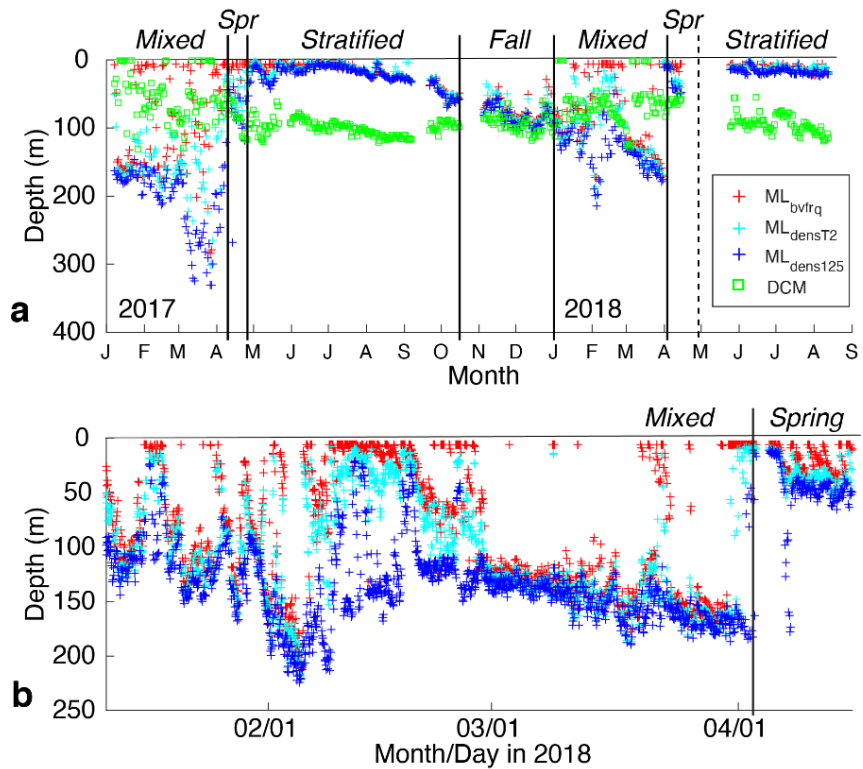

**Figure S2. Comparison of MLD definitions from glider profiles.** **(a)** Daily average MLD and DCM computed for the 2017 to 2018 period, colored symbols are defined in the legend. Vertical lines depict season transitions; the end of *Spring* 2018 (dashed) occurred during a period of interrupted observations. **(b)** MLD computed in individual profiles for a single glider mission 12 January - 15 April 2018.